\hsize=126mm

\vsize=195mm

\def\frac#1#2{{\begingroup#1\endgroup\over#2}}

\centerline{\bf A NOTE ON THE MILKY WAY AS A BARRED GALAXY}
\bigskip

\centerline{S. SAMUROVI{\'C}$^{1*}$, V. MILO{\v S}EVI{\'C}--ZDJELAR,$^{1**}$
 M.  M. {\'C}IRKOVI{\'C}$^{2,3}$ and} 

\centerline{V. {\v C}ELEBONOVI{\'C}$^4$}

\bigskip

\centerline{\it $^1$ Public Observatory, Gornji Grad 16, Kalemegdan, 11000 
Belgrade, Yugoslavia} 

\centerline{\it $^*$E-mail srdjanss@afrodita.rcub.bg.ac.yu}

%\centerline{\it $^2$ Public Observatory, Gornji Grad 16, Kalemegdan, 11000 
%Belgrade, Yugoslavia} 

\centerline{\it $^{**}$ E-mail vesnamz@afrodita.rcub.bg.ac.yu}

\medskip

\centerline{\it $^2$ Dept of Physics \& Astronomy, SUNY at Stony Brook,}

\centerline{\it Stony Brook, NY 11794-3800, USA}

\centerline{\it $^3$ Astronomical Observatory, Volgina 7, 11000 Belgrade, 
Yugoslavia} 

\centerline{\it E-mail  cirkovic@sbast3.ess.sunysb.edu}

\medskip

\centerline{\it $^4$ Institute of Physics, Pregrevica 118, 11080 Zemun-
Belgrade, Yugoslavia} 

\centerline{\it E-mail celebonovic@exp.phy.bg.ac.yu} \bigskip 
\noindent{\bf 
Abstract:} We review  recent research on the Milky Way galaxy and try to 
investigate whether its shape is similar to other barred galaxies. The 
emphasis is given on  microlensing research because this method can be 
useful in determining the shape of the Galaxy with the minimal set of
assumptions. By analyzing  plots of the microlensing optical depth, $\tau$ 
as a function of galactic coordinates  for different values 
of the axis ratio, $q$ of the galactic halo, we have shown  that observations
 are best described by a flattened halo  with  $0.2^<_\sim$ $q$ ${}^<_\sim 0.6$. 

\bigskip

\noindent{\bf Introduction}
\medskip

Over the last decade a growing list of evidence led us to accept the 
hypothesis proposed back in 1964 by  de Vaucouleurs (e.g. 
Kuijken 1996): our Galaxy is a barred one. The arguments were 
initially based upon morphological considerations.  The development 
of observational techniques broadens the possibilities  of investigation of the 
central parts of the Milky Way in several research fields. All of them, 
however, confirm de Vaucouleurs' idea that the Milky Way is a SAB(rs) type  
galaxy, having a bar, weak rings, and a four-arms spiral structure (Vall\'ee 
1995).   

We shall only mention here the following 
observational results (Kuijken 1996)  that confirm these facts: 

\item {1.} {\bf photometric research} that include:

\item \item {a)} surface photometry (Blitz and  Spergel 1991; Dwek {\it et 
al.\/} 1995); and

\item \item {b)} star counts (Gould, Bahcall and Flynn 1997; Nikolaev 
and Weinberg 1997) 

\item {2.} {\bf kinematical research} that include:

\item \item {a)} gas kinematics   (Binney {\it et 
al.\/} 1991); and 

\item \item {b)} stellar kinematics  (Kuijken 1996, and references therein)

\item {3.} {\bf gravitational microlensing (ML) researches}.

\medskip

The existence of the bar successfully explains dynamical peculiarities observed 
in our Galaxy, and shows that they are normal features in a barred galaxy: 

-- stellar and gas dynamics in the Galactic center region (Binney {\it et al.}
 1991),

--   central activity definitely correlated with a presence of the bar in 
galaxies (Combes {\it et al.\/} 1995),

-- a $2.6\times 10^6\, M_\odot$ black 
hole in the center of the Milky Way  (Bower and  Backer 1998; Falcke {\it et 
al.} 1998), 

-- the absence of HI and CO gas in the region $1.5\; {\rm kpc} \le r \le 3.5\;
 {\rm kpc}$, implying that the major axis of the bar is $r_{bar}\le r_{cr}$,  
where  $r_{cr}=2.4$ kpc is the Milky Way's corotation radius (Binney and Tremaine 
1987; Binney {\it et al.} 1991), 

-- a molecular ring at 3.5 kpc, which can  be explained by  gas accumulation 
near the outer Lindblad resonance radius ($r_{\rm OLR}=4.1$ kpc in Milky Way) 
(Binney {\it et al.\/} 1991; Freundreich 1998), 

-- asymmetry in  the  distribution of the  red clump stars  in the bulge (OGLE)
 (Stanek 1995),

-- asymmetries of the bulge photometric image (COBE-DIRBE) (Dwek 1995, Binney, 
Gerhard and Spergel 1996), and

-- excess of gravitational microlensing  events compared to the theoretical 
estimates (Paczy\'nski {\it et al.} 1994) in the galactic bulge direction.

All observations agree that the value of the bar inclination angle (to the 
Sun-Galactic Centre line) is between  $10^ \circ$ and $30^ \circ$.

In this paper we will   discuss  gravitational 
ML research and try to establish the connection between the 
inner parts of our Galaxy (bar) and its outer parts (halo) 
using this new observational technique. The ultimate goal is 
to see how the Milky Way as 
barred galaxy can be compared to  other barred spirals and whether some 
conclusions concerning the shape of the Galaxy can be drawn. \bigskip

\noindent{\bf Baryonic dark matter content of the Galaxy}
\medskip

The  dark matter (DM) content of the Milky Way is still unknown. From 
shape of its rotation curve (RC) (Merrifield 1992) one can see that a huge 
amount of mass still has to be identified.  The difficulties 
in the determination of the RC led to  uncertainties in the most 
important parameters such as the galactic constant $R_0$, which represents the 
distance to the Galactic center and  the circular speed at the Solar 
radius, $v_0$ (Merrifield 1992, Olling and Merrifield 1998, Sackett 1997).
 Although 
the IAU 1986 standard values are: $R_0=8.5$ kpc and $v_0=220\, {\rm km s^{-
1}}$ some recent estimates allow the smaller values: $R_0=7.1\pm 0.4$ kpc and 
$v_0=184\pm 8\, {\rm km s^{-1}}$ (Olling and Merrifield 1998). In this paper 
we adopt the value $R_0=8.5\pm 
0.5$ kpc (Feast and Whitelock 1997) based upon an analysis of Hipparcos proper 
motion of 220 Galactic Cepheids and $v_0=210\pm 25\, {\rm km s^{-1}}$  that 
includes the best values from the HI analysis ($v_0=185\, {\rm km s^{-1}}$) 
and the estimated values  based on the Sgr A* proper motion ($v_0=235\, 
{\rm km s^{-1}}$) (Sackett 1997).

Without going into the  discussions about the content of the DM in the halo, we
only state here that one part (presumably smaller) has to be in the baryonic 
form. Namely, cosmic nucleosynthesis predicts that (Turner 1996): 

$$0.008\;^<_\sim \;\Omega_B\; h^2\; ^<_\sim \;0.024\eqno (1)$$ where 
$\Omega_B$ is the universal baryonic mass-density parameter ($\Omega_B=8\pi 
G\rho _B/3H_0^2$) and  $ 0.4^< _\sim\; {h}\;^<_\sim 1.0$. ``Silent'' $h$ is 
used in parametrization of the Hubble constant $H_0=100h\; {\rm km\,s^{-
1}\,Mpc^{-1}}$. Recent estimates (Fukugita, Hogan and Peebles 1998) give: 
 $$0.007\;^<_\sim \;\Omega_B\;  ^<_\sim \;0.041\eqno (2)$$ 
with the ``best guess'' $\Omega _B\sim 0.021$ (for $H_0=70 \; {\rm km\,s^{-
1}\,Mpc^{-1}}$). 

Using the simplest dynamical estimate of the mass of the Galaxy (Kepler's 
third  law): $$G M(r)=v(r)^2\eqno(3)$$ where $M(r)$ is the mass interior to 
$r$, $v$ is the measured rotational velocity and $r$ is the radius within 
which most of the light in galaxy is emitted.  For luminous 
matter one can obtain: 
$$0.003^<_\sim \; \Omega_{\rm LUM}\; ^<_\sim 0.007\eqno(4)$$
(Roulet and Mollerach 1997, and references therein).
This is consistent with severe limits on mass-to-light ratio in the Local
Group imposed by deep blank sky surveys (Richstone {\it et al.} 1992; Hu {\it
et al.} 
1994; Flynn, Gould and Bahcall 1996), as well as
with huge dynamical mass for the Milky Way inferred by Kulessa and
Lynden-Bell (1992).

 The mass in the halo is dominated by the matter that is not, at least easily, 
detectable. So, one can write: 
$$\Omega_{\rm HALO}\,{}^>_\sim 0.1 \; {}^>_\sim 14\, \Omega_{\rm LUM}.\eqno(5)$$

It can be seen the equations (2) and (4)  that  dark 
baryonic matter must exist; various types of such   material have been suggested: 
gaseous clouds of plasma or neutral atoms and molecules, snowballs or icy 
bodies similar to comets, stars, planets, white dwarfs, neutron stars and 
stellar or primordial black holes (e.g. Peebles 1993). \bigskip 

\noindent{\bf Microlensing -- methods and results}
\medskip

In searches for the baryonic DM content the method of microlensing has so far 
proved successful. Its name derives from the fact that  lensing of distant 
objects is made by bodies with masses characteristic of a star or planet. 
Although the theoretical development of this idea started in 1964 (e.g. 
Peebles 1993, and references therein), it was the seminal paper by Paczy\'nski  
(1986) that showed that one can search for ML events in the Milky Way halo if 
it is made of stars or brown dwarfs. Rapid development of observational and 
computer technology led to the detection of  a significant number of ML events 
(e.g. Mellier, Bernardeau and Van Waerbeke 1998). The directions include Large 
and Small Magellanic Clouds (LMC and SMC) (Alcock {\it et al.} 1996, 1997b; 
Palanque-Delabrouille {\it et al.} 1998), Galactic bulge (Kiraga and Paczy\'nski 
1994) and M31 (Crotts 1996).

All these surveys give results concerning two important parameters: masses 
of the intervening objects and the optical depth. In the Table 1 we give the 
targets observed, names of the appropriate survey, mass ranges of the 
lenses, and corresponding optical depth.  

\vfill\eject

\settabs=4\columns
\+ Target & Survey & Mass range & Optical depth\cr
\smallskip
\hrule
\smallskip
%\+ {\rm LMC/SMC} & {\sevenrm EROS1} & $\approx$ 0.15 $M_\odot$& &\cr
\+ {\rm LMC/SMC} & {\sevenrm MACHO} & $\approx$ 0.3 -- 0.5 $M_\odot$ 
& $\tau_{\rm LMC}=2.9^{+1.4}_{-0.9}\times 10^{-7}$\cr
\+ & & &$\tau_{\rm SMC}=1.5-3\times 10^{-7}$\cr
\+ Gal. bulge & {\sevenrm MACHO;DUO;OGLE}&0.08 -- 0.6 $M_\odot$ & 
$\tau_{\rm bul}=3.9^{+1.8}_{-1.2}\times 10^{-6}$&\cr  
\+ {\rm M31} & {\sevenrm KPNO} & $\approx$ 10 $M_\odot$ & $\tau_{\rm M31}
=5-10\times 10^{-6}$ &\cr
\+ {\rm LMC/SMC} & {\sevenrm EROS2} & 0.85 -- 8.7 $M_\odot$ &
 $\tau _{\rm SMC}=3.3\times 10^{-7}$&\cr
\smallskip
\hrule
\medskip

\noindent {\bf Table 1}. Targets in different ML surveys, the mass 
ranges of the lenses and optical depths. 

\medskip

 Another important quantity, the optical depth, $\tau$ is used in 
discussion of ML and as  we shall show, in determining the shape of the 
halo. It can be defined as the probability that at a given time a source 
star is being microlensed with an amplification larger than 1.34 (e.g. Roulet 
and Mollerach 1997). 

Here we wish to investigate in more details one property of the halo of the 
Milky Way that has often been  neglected: its shape. It is known from 
the work of Sackett and Gould (1993) that  instead  of the equation 
for the mass density in a   spherical halo: 

$$\rho ({\bf r})={v_\infty^2 \over 4\pi G}\left ({1\over a^2+r ^2}\right 
)\theta (R_T-r)\eqno (6)$$
 (where $r$ is the Galactocentric radius, $v_\infty$ 
is the asymptotic circular speed of the halo, $a$ is the core radius of the 
halo and $R_T$ is the truncation radius) one should use the generalized formula: 
$$\rho ({\bf 
r})={\tan \psi\over \psi}{v_\infty^2 \over 4\pi G}\left ({1\over a^2+\zeta 
^2}\right )\theta (R_T-\zeta)\eqno (7)$$
 where $\zeta ^2=r^2+z^2\tan^2\psi$ ($z$ denotes height above the Galactic 
plane). Here the flattening parameter $\psi$ is introduced: $\cos \psi=q=c/a$, 
i.e. its cosine is equal to the axis ratio and determines the shape of the 
halo $En$. $En$ is related to $q$ as $q=1-n/10$. Following Sackett and Gould 
(1993) we write the following expression for the estimate of the optical depth 
as a function of Galactic coordinates $l$ (longitude) and $b$ (latitude): 
$$\tau(l,b)={\tan\psi\over \psi} {v_\infty^2 \over c^2}{1\over D}\int _0 
^{D}{dL(D-L)L\over (a^2+R_0^2)-(2R_0\cos l \cos b)L 
+(1+\sin^2b\tan^2\psi)L^2}\eqno(8)$$ where we put $R_0=8.5$ kpc and $a=5$ 
(e.g. Alcock 1996). Now we integrate this equation and take $D=50$ kpc (for 
LMC), $D=63$ kpc (for SMC) and $D=770$ kpc for M31. Although Sackett and Gould  
(1993) take values for $q$ starting with $q=0.4$ (shape E6) we will start with 
admittedly extreme value $q=0.2$ (shape E8) required by some theories such as 
DDM (decaying dark matter) theory (Sciama 1997), based upon the recent 
Dehnen-Binney models of the Galaxy (Dehnen and Binney 1998). Attempts were made to 
show that this small value of $q$ is not possible since $q=0.75\pm 0.25$ 
(Olling and Merrifield 1997), but at the cost that $R_0=7.1\pm 0.4$ kpc 
(Olling and Merrifield 1998). We will nevertheless take into account such 
small value for $q$ since we find DDM theory acceptable in solving  different 
serious astrophysical and cosmological problems (e.g. Sciama 1993). 

There are several other lines of reasoning suggesting a high degree of halo 
flattening in spiral galaxies. One is for long time suspected (e.g. Ninkovi\'c 
1985) flattening of the Population II subsystem, which may be a consequence of 
the residual rotation, or more probably, global flattening of the 
gravitational potential created by dark matter. The other is the  behavior of 
the gas distributed in the halo. If the seminal idea of Bahcall and Spitzer 
(1969) of extended gaseous halos of normal galaxies producing narrow 
absorption features in the spectra of background objects is correct, as 
indicated by recent low-redshift measurements (Bergeron and Boiss\'e 1991; 
Lanzetta {\it et al.} 1995), then the distribution of gas could tell us 
something about the shape of the gravitational potential. It is not a simple 
problem at all (see  Barcons and Fabian 1987), but some results are quite 
suggestive. In an important recent paper, Rauch and Haehnelt (1995) have shown 
that for the most plausible values of Ly$\alpha$ cloud parameters, the 
conclusion that their axial ratio (thickness/transverse length) is less than 
0.25 is inescapable. This conclusion does not depend on the exact choice of 
model for Ly$\alpha$ clouds, and, if the observations quoted above are 
correctly interpreted, would mean that the gaseous halos are also flattened by 
the same amount. One should keep in mind, though, that such absorption studies 
probe only ``a tip of an iceberg'', since these objects are ionized to 
extremely high degree, and may as well contain dominant part of the baryonic 
density in eq.~(1). 

Bearing this in mind,  we solve the integral in the eq.~(8)  and give estimate 
for $\tau$ in several cases of particular interest: 

\item {$\bullet$} Optical depth $\tau (l,b)$ in the parametric space, with the 
parameter $q$ fixed in steps of 0.2, i.e. $q=0.2$, $q=0.4$, $q=0.6$, $q=0.8$ 
and $q\approx 1$. 

\item {$\bullet$} Optical depth $\tau$ for different targets: LMC, SMC, M31 
and Galactic bulge (bar) in order to see what value of $q$ determines the 
optical depth that is closest to observed value in the appropriate survey. 

Due to the space limitations, we hereby present just two three-dimensional 
plots. In the Figure 1, value of the optical depth $\tau$  is plotted against 
galactic coordinates $l$ and $b$. This is an estimate for $q=0.4$, but 
 can  easily be done for other values. One can use such plots (on the same  or  
smaller angular scales) in order to choose observing direction where the 
optical depth reaches maximal values. Such an example is shown in the Figure 2 
where we plotted optical depth as function of coordinates $l=280.^\circ 5$ and 
$b=-32.^\circ 9$  of the Large Magellanic Cloud. Other plots and results of 
integration will be presented elsewhere.\footnote{$^1$}{Some examples of  
plots and  results of integration can be found in the postscript format at the  
following {\tt URL}: {\tt http://www.geocities.com/CapeCanaveral/7102/ 
 Belgrade-MACHO.html}, or from the authors by e-mail.} 

 \medskip

\noindent{\bf Results and conclusions}

After solving the integral in the eq.~(8) for given values of the parameter  
$q$ we looked for the values that match the optical depth obtained in various 
surveys. We found that the best agreement can be attained if we take 
$0.2^<_\sim$ $q$ ${}^<_\sim 0.6$. Namely: 

\item {1.} For the case of the  LMC, that has been studied rather well, the 
measured value of the optical depth based upon the sample of 8 events is 
$\tau=2.9^{+1.4}_{-0.9}\times 10^{-7}$ (Alcock 1997b) while we find that for 
$q=0.5$ we have $\tau\approx 3\times 10^{-7}$  (see Figure 2). 

\item {2.} For the case of the SMC, that is studied less thoroughly, the 
optical depth is estimated as $\tau=1.5-3\times 10^{-7}$ (Alcock 1997c).   Our 
results show that the model in the eq.~(8) gives the value $\tau ^>_\sim 
4\times 10^{-7}$ for $q\approx 0.5$ and above. 

\item {3.} For the case of the galaxy M31 we found $\tau\approx 5\times 10^{-
6}$ which is an accordance with the estimates $5-10\times 10^{-6}$ (Crotts 
1996), under the assumption that $q^>_\sim 0.2$. 

\item{4.} Determining $\tau$ towards the Galactic center is more  complicated 
and we will not discuss it here. We only state that using the model in the 
eq.~(8) we can estimate the halo contribution to the ML rate towards Galactic 
center which is between $\tau\approx 5\times 10^{-8}$ ($q=0.6$) and 
 $\tau\approx 1.6\times 10^{-7}$ ($q=0.2$); the estimated range for 
the total optical depth towards Galactic center is 
($\tau=3.9^{+1.8}_{-1.2}\times 10^{-6}$) (Alcock 1997a).

{}From our estimates it is obvious that the spherical dark halo can be ruled 
out: the value for $q$ lies in the  interval: $0.2^<_\sim$ $q$ ${}^<_\sim 
0.6$. 
Recent research shows that it is not uncommon case with spiral galaxies 
(Sackett and Sparke 1990; Sackett {\it et al.} 1994; Olling 1995). 
Very recently, 
 the observations of the gravitational lens system B1600$+$434, 
consisting of two spiral galaxies (G1 and G2), where G2 is a barred one, 
suggest that it has 
 $q^>_\sim 0.4$ (Koopmans, de Bruyn and Jackson 1998).

\bigskip

\centerline{\bf References}
\medskip

\item{}\kern-\parindent{Alcock, C. {\it et al.} (MACHO collaboration): 1996, 
{\it Astrophys. J.}, {\bf 461}, 84.}

\item{}\kern-\parindent{Alcock, C. {\it et al.} (MACHO collaboration): 1997a, 
{\it Astrophys. J.}, {\bf 479}, 119.}

\item{}\kern-\parindent{Alcock, C. {\it et al.} (MACHO collaboration): 1997b, 
{\it Astrophys. J.}, {\bf 486}, 697.} 

\item{}\kern-\parindent{Alcock, C. {\it et al.} (MACHO collaboration): 1997c, 
{\it Astrophys. J.}, {\bf 491}, L11.}

%Bally, J., Stark, A.A., Wilson, R.W. and Henkel, C.: 1988, {\it Astrophys. J.}, 
{\bf 324}, 223.

\item{}\kern-\parindent{Bahcall, J.N. and Spitzer, L.: 1969, {\it Astrophys. J.}, 
{\bf 156},  L63.}

\item{}\kern-\parindent{Barcons, X. and Fabian, A. C.: 1987, {\it MNRAS}, {\bf 
224}, 675.} 

\item{}\kern-\parindent{Bergeron, J. and Boiss\'e, P. 1991, {\it Astron.
Astrophys.}, {\bf 
243}, 344.} 

\item{}\kern-\parindent{Binney, J. and Tremaine, S.: 1987, {\it Galactic 
Dynamics}, Princeton University Press, Princeton.}

\item{}\kern-\parindent{Binney, J., Gerhard, O.E., Stark, A.A., Bally, J. and 
Uchida, K.I.: 1991, {\it MNRAS}, {\bf 252}, 210.}

\item{}\kern-\parindent{Binney, J., Gerhard, O.E and Spergel, D.: 1997, {\it 
MNRAS}, {\bf 288}, 365.}

\item{}\kern-\parindent{Blitz, L. and Spergel, D.N.: 1991, {\it Astrophys. 
J.}, {\bf 379}, 631.}

\item{}\kern-\parindent{Bower, G.C. and  Backer, D.C.: 1998, preprint 
astro-ph/9802030.}

\item{}\kern-\parindent{Combes, F., Boiss\'e, P., Mazure, A. and Blanchard, 
A.: 1995, {\it Galaxies and Cosmology}, Springer, Berlin.

\item{}\kern-\parindent{Crotts, A.P.S.: 1996, paper  presented at the International 
Conference on ``Dark and Visible Matter in Galaxies'' (Sesto Pusteria, Italy, 
2-5 July 1996) (preprint astro-ph/9610067).}

\item{}\kern-\parindent{Dehnen, W. and Binney, J.: 1998, {\it MNRAS}, {\bf 
294}, 429.}

\item{}\kern-\parindent{Dwek, E., Arendt, R.G., Hauser, M.G., Kelsall, T., 
Lisse, C.M., Moseley, S.H., Silverberg, R.F., Sodroski, T.J. and Weiland, 
J.L.: 1995, {\it Astrophys. J.}, {\bf 445}, 716.} 

\item{}\kern-\parindent{Falcke, H., Goss, W.M., Matsuo, H., Teuben, P., Zhao, J.H. 
and 
Zylka, R.:  1998, {\it Astrophys. J.} in press, (preprint astro-ph/9801085).}

\item{}\kern-\parindent{Feast, M. and Whitelock, P.: 1997, {\it MNRAS}, {\bf 291},
 683. }

\item{}\kern-\parindent{Flynn, C., Gould, A. and  Bahcall, J. N.: 1996, {\it 
Astrophys. J.}, {\bf 466}, L55.} 

\item{}\kern-\parindent{Freudenreich, H.T.: 1998, {\it Astroph. J.}, {\bf 
492}, 495.}

\item{}\kern-\parindent{Fukugita, M., Hogan, C.J. and Peebles, P.J.E.: 1998, 
{\it Astrophys. J.}, submitted (preprint astro-ph/9712020).} 

\item{}\kern-\parindent{Gould, A., Bahcall, J.N. and Flynn, C.: 1997, 
{\it Astrophys. J.}, {\bf 482}, 913..}

%\item{}\kern-\parindent{Griest, K.: 1991, {\it Astrophys. J.}, {\bf 366}, 
%412.} 

\item{}\kern-\parindent{Hu, E. M., Huang, J.-S., Gilmore, G. and Cowie, L. L.: 
1994, {\it Nature}, {\bf 371}, 493.}

%\item{}\kern-\parindent{Izumiura, H., Deguchi, S. and Fujii. T.: 1998,
 {\it Astrophys. J. Lett.\/} in press (preprint astro-ph/9712101)}

\item{}\kern-\parindent{Kiraga, M. and Paczy\'nski, B.: 1994, {\it Astrophys. 
J.}, {\bf 430}, L101.}

\item{}\kern-\parindent{Kulessa, A. and Lynden-Bell, D.: 1992, {\it MNRAS}, 
{\bf 255}, 105.} 

\item{}\kern-\parindent{Koopmans, L.V.E., de Bruyn, A.G. and Jackson, N.: 
1998, {\it MNRAS} accepted (preprint astro-ph/9801186).} 

\item{}\kern-\parindent{Lanzetta, K. M., Bowen, D., Tytler, D. and Webb, J. 
K.: 1995, {\it Astrophys. J.}, {\bf 442}, 538.

\item{}\kern-\parindent{Kuijken, K.: 1996, in {\it IAU Coll. 157}, eds. R. 
Buta,  D.A. Crocker, B.G. Elmegreen, ASP, San Francisco, 504.}

\item{}\kern-\parindent{Mellier, Y., Bernardeau, F. and Van Waerbeke, 
L.: 1998, preprint astro-ph/ 9802005.}

\item{}\kern-\parindent{Merrifield, M.M.: 1992, {\it Astron. J.}, {\bf 103},  
1552.}

\item{}\kern-\parindent{Nikolaev, S. and Weinberg, M.D.: 1997, {\it Astrophys. 
J.}, {\bf 487}, 885..}

\item{}\kern-\parindent{Ninkovi\'c, S.: 1985, {\it Ap \& SS}, {\bf 110}, 379.}

\item{}\kern-\parindent{Olling. R.P.: 1995, Ph.D. thesis, Columbia University.}

\item{}\kern-\parindent{Olling, R.P. and Merrifield, M.R.: 1997, talk 
presented at Workshop on Galactic Halos, UC Santa Cruz, August 1997 (preprint 
astro-ph/9710224).}

\item{}\kern-\parindent{Olling, R.P. and Merrifield, M.R.: 1998, {\it MNRAS} 
accepted (preprint astro-ph/\break 9802034).}

\item{}\kern-\parindent{Paczy\'nski B.: 1986, {\it Astrophys. J.}, {\bf 304}, 
1.} 

\item{}\kern-\parindent{Paczy\'nski B. {\it et al.} (OGLE Collaboration): 1994, 
{\it Astrophys. J.}, {\bf 435}, 
L11.}

\item{}\kern-\parindent{Palanque-Delabrouille, N. {\it et al.}
 (EROS Collaboration): 1997, preprint astro-ph/ 9710194.}

\item{}\kern-\parindent{Peebles, P.J.E.: 1993, {\it Principles of
Physical Cosmology}, Princeton University Press, Princeton.

\item{}\kern-\parindent{Rauch, M. and Haehnelt, M. G.: 1995, {\it MNRAS}, {\bf 
275}, L76.}

\item{}\kern-\parindent{Richstone, D., Gould, A., Guhathakurta, P. and Flynn, 
C.: 1992, {\it Astrophys. J.}, {\bf 388}, 354.}

\item{}\kern-\parindent{Roulet, E. and Mollerach, S.: 1997, {\it Phys. 
Rep.}, {\bf 279}, 68.}

\item{}\kern-\parindent{Sackett, P.D. and Sparke, L.S: 1990, 
{\it Astrophys. J.}, {\bf 361}, 408.}

\item{}\kern-\parindent{Sackett, P.D. and Gould, A.: 1993, {\it Astrophys. 
J.}, {\bf 419}, 648.}

\item{}\kern-\parindent{Sackett, P.D., Rix, H.-W., Jarvis, B.J., 
Freeman, K.C.: 1994, {\it Astrophys. 
J.}, {\bf 436}, 629.}

\item{}\kern-\parindent{Sackett, P.D.: 1997, {\it Astrophys. J.}, {\bf 483}, 
103.}

\item{}\kern-\parindent{Sciama, D.W.: 1993, {\it Modern Cosmology and the Dark 
Matter Problem}, Cambridge University Press, Cambridge.} 

\item{}\kern-\parindent{Sciama, D.W.: 1997, preprint astro-ph/9704081.}

\item{}\kern-\parindent{Stanek, K.Z.: 1995, {\it Astrophys. J.}, {\bf 441}, L29.}

\item{}\kern-\parindent{Turner, M.S.: 1996, preprint astro-ph/9610158.}

\item{}\kern-\parindent{Val\'ee, J.P.: 1995, {\it Astrophys. J.}, {\bf 454}, 
119.} 

\bigskip
\bigskip
\noindent Fig. 1. Optical depth $\tau$ as a function of the galactic coordinates
 $l$ and $b$. Distance to  sources is taken to be 50 kpc and the axes ratio is
  $q=0.4$.             

\bigskip

\noindent Fig. 2. Optical depth $\tau$ as a function of the galactic coordinates 
$l$ and $b$ for the Large Magellanic Cloud (LMC) for the axis ratio $q=0.5$.

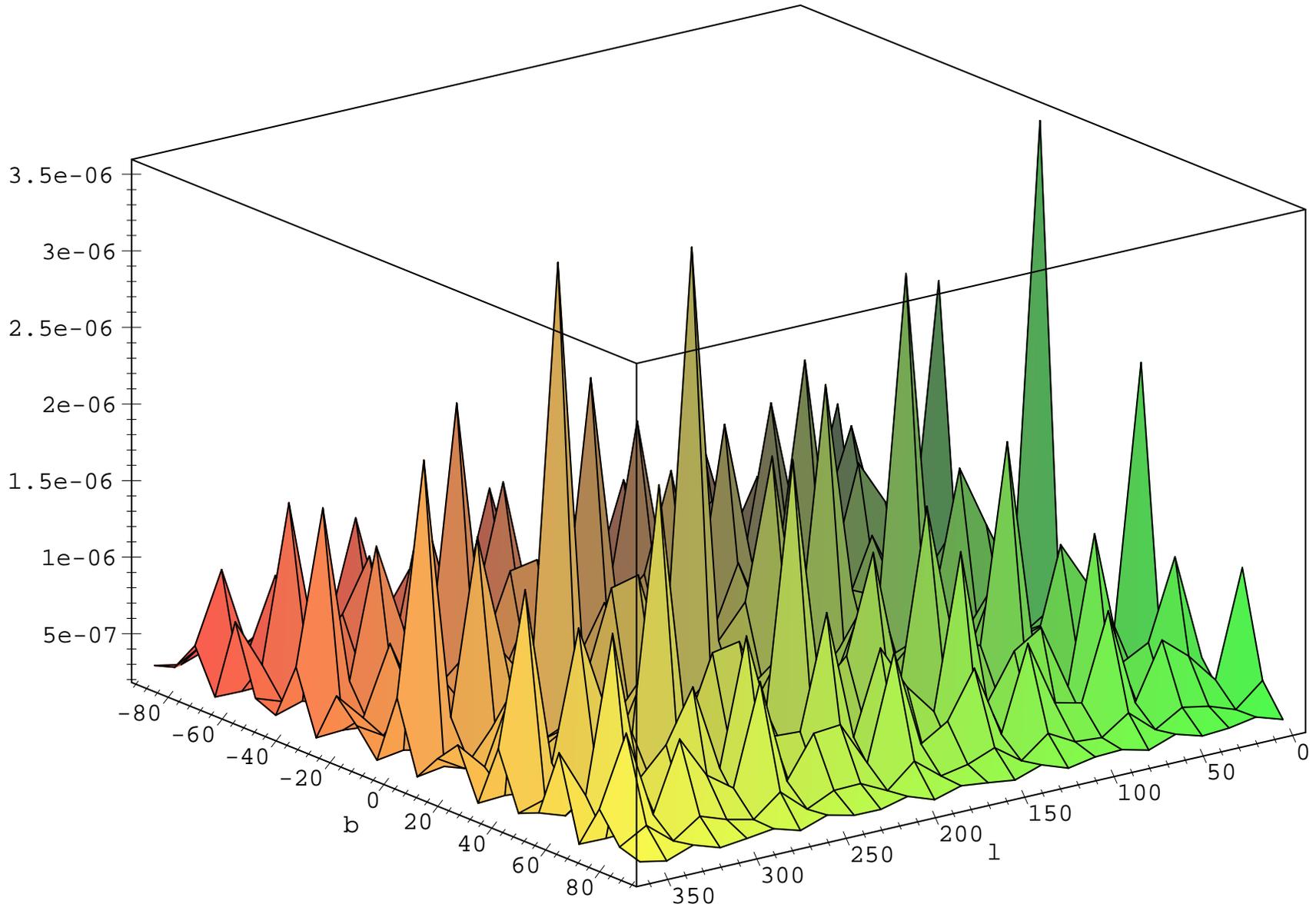
\bye